# Effects of van der Waals interactions and quasiparticle corrections on the electronic and transport properties of Bi$_2$Te$_3$

L. Cheng[1], H. J. Liu[1,*], J. Zhang[1], J. Wei[1], J. H. Liang[1], J. Shi[1], X. F. Tang[2]

[1]*Key Laboratory of Artificial Micro- and Nano-Structures of Ministry of Education and School of Physics and Technology, Wuhan University, Wuhan 430072, China*

[2]*State Key Laboratory of Advanced Technology for Materials Synthesis and Processing, Wuhan University of Technology, Wuhan 430070, China*

We present a theoretical study of the structural, electronic and transport properties of bulk Bi$_2$Te$_3$ within density functional theory taking into account the van der Waals interactions (vdW) and the quasiparticle self-energy corrections. It is found that the optB86b-vdW functional can well reproduce the experimental lattice constants and interlayer distances for Bi$_2$Te$_3$. Based on the fully optimized structure, the band structure of Bi$_2$Te$_3$ is obtained from first-principles calculations with the GW approximation and the Wannier function interpolation method. The global band extrema are found to be off the high-symmetry lines, and the real energy band calculated is in good agreement with that measured experimentally. In combination with the Boltzmann theory, the GW calculations also give accurate prediction of the transport properties, and the calculated thermoelectric coefficients of Bi$_2$Te$_3$ almost coincide with the experimental data.

## I. INTRODUCTION

As a traditional good thermoelectric material, bismuth telluride has attracted lots of experimental and theoretical investigations on the electronic and transport properties. Bulk Bi$_2$Te$_3$ has a rhombohedral structure belongs to the space group $R\bar{3}m$ with five atoms in the unit cell (Fig. 1). They form stacks of covalently bonded quintuple layers (QLs) in the sequence of Te1-Bi-Te2-Bi-Te1, where the QLs are bounded through weak van der Waals (vdW) interactions. Unfortunately, the traditional density functional theory (DFT) is unable to account for long range vdW interactions.

---

* Author to whom correspondence should be addressed. Electronic mail: phlhj@whu.edu.cn.

Although vdW interaction is usually very weak, it can have a significant effect on the inter-QL distance [1] which in turn may play an important role in determining the electronic and transport properties of bulk $Bi_2Te_3$. Recently, Dion *et al.* [2] proposed a non-local correlation functional which approximately accounts for the dispersion interactions. The so-called vdW-DF is implemented into DFT to deal with the weak vdW interactions [3]. Cooper reported an exchange functional that is compatible with the Rutgers-Chalmers vdW correlation functional [4]. Luo *et al.* [1] investigated the atomic, electronic, and thermoelectric properties of bulk $Bi_2Te_3$ and $Bi_2Se_3$ with the vdW interactions included, and they found that using the vdW-DF$^{C09}_x$ functional gives a much better agreement with the experimental structural parameters. Moreover, their calculations indicated that vdW interactions are crucial for the electronic and thermoelectric properties. Liu *et al.* [5] considered few QLs films of $Bi_2Te_3$ and $Bi_2Se_3$, and found that the vdW interactions are needed to obtain correct lattice parameters and energy band structures.

Recently, $Bi_2Te_3$ has been subject to extensive study not only as a good thermoelectric material, but also as a good topological insulator (TI) [6, 7, 8]. The theoretical study of TIs involves very precise electronic calculations for the excited-state properties, which is beyond the scope of traditional DFT. One approach to overcome this deficiency is to calculate the quasiparticle properties with the GW approximation of the many-body effects [9, 10]. In fact, the GW method has been applied to study the electronic structures of $Bi_2Te_3$ in several previous reports [11, 12, 13, 14]. In Ref. 11, Kioupakis *et al.* calculated the quasiparticle band structures of $Bi_2Te_3$ with the experimental lattice parameter and found that it has direct and indirect band gaps that are very close to each other (~0.17 eV). In Ref. 13, Nechaev and Chulkov showed how the atomic positions, the DFT exchange correlation founctionals, and the GW approximation affect the location of band extrema and thus the band gap of $Bi_2Te_3$. They found that the GGA+GW with relaxed atomic positions gives a band gap of 156 meV, which agrees well with the experimental value. The conduction band minimum (CBM) and the valance band maximum (VBM) they found is similar to those appearing in angle resolved photoemission spectroscopy (ARPES) measurements [15].

Although extensive electronic structures and transport calculations have been performed for $Bi_2Te_3$, however, most of them are calculated using the experimental structural parameters without fully relaxations in a self-consistent way. There are also theoretical works taking into account the vdW interactions [1, 5], or including the GW approximation [11, 12, 13, 14]. However, to the best of our knowledge, no work has been reported so far considering both of them, especially for their combined effects on the electronic and transport properties of $Bi_2Te_3$. In the present work, we first optimize the atomic structure of $Bi_2Te_3$ by using appropriate vdW functionals. We next calculate the electronic structure of $Bi_2Te_3$ with the vdW interactions and GW approximation both considered. The transport properties are then derived by a combination of Boltzmann theory and Wannier interpolation method. The thermoelectric transport properties of $Bi_2Te_3$ predicted from our theoretical calculations are in excellent agreement with those measured experimentally.

## II. COMPUTATIONAL DETIALS

Our theoretical calculations have been performed by using the projector augmented-wave method [16, 17] within the framework of DFT [18, 19, 20]. The exchange correlation energy is in the form of Perdew-Burke-Ernzerhof [21] with generalized gradient approximation (GGA). For structural optimization, the energy cutoff is set as 400 eV and the Brillouin zone is sampled with 9×9×9 Monkhorst-Pack $k$ meshes. Optima atom positions are determined until the magnitude of the force acting on each Bi atom becomes less than 0.01 eV/Å. In order to illustrate the vdW interactions effects on the lattice parameters, a detailed investigation with different forms of vdW exchange correlation functionals have been performed. In the GW calculations, the Brillouin zone is sampled with 6×6×6 Γ centered grids, and the number of unoccupied bands is set as 272. The electronic structures are then obtained using the Wannier interpolation formalism [22, 23, 24]. To locate the exact VBM and CBM and thus the real band gap, a very dense $k$ meshes up to 4960 points is used in the irreducible Brillouin zone. The electronic transport properties are computed using the semiclassical Boltzmann theory with the maximally-localized Wannier function basis to interpolate band structures and band velocities [25]. The spin-orbit coupling is explicitly considered in our calculations.

## III. RESULTS AND DISCUSSIONS

**A. Structural properties of $Bi_2Te_3$ including vdW intreractions**

The crystal structure of $Bi_2Te_3$ is shown in Fig. 1. The experimentally measured lattice parameters are $a_0$=10.476 Å and $\alpha$=24.166° [26], which corresponds to $a$=4.386 Å and $c$=30.497 Å in a hexagonal cell. The internal coordinates are $u$=0.4000 for Bi atom and $v$=0.2095 for Te1 atom, and the inter-QL distance is $d_{eqm}$=2.612 Å. On the theoretical side, extensive DFT calculations [27, 28, 29, 30, 31, 32] have been performed for bulk $Bi_2Te_3$. However, most of them deal with the experimental lattice parameters, and their calculations do not properly treat the long-range dispersion interactions. There is currently growing interest in the calculations with vdW interactions explicitly implemented, and various kinds of vdW functionals have been suggested [2, 3]. Using these functionals, we calculate the structural parameters of $Bi_2Te_3$ and our results are summarized in Table I. The experiment results are also given for comparison. It is obvious that the standard DFT with PBE functional tends to overestimate the lattice parameters. Such overestimation will be reduced more or less when the vdW interactions are considered in the calculations. This observation holds for all the mentioned vdW functionals except for the vdW-DF2-rPW86 functional. It should be mentioned that $a_0$ and $c$ calculated with vdW-DF2-rPW86 are very close to that predicted with vdW-DF$^{revPBE}_x$ functional [1]. For all the investigated functionals, the difference in the internal coordinates is very small, which is consistent with that found in Ref. 1. As for the inter-QL distance $d_{eqm}$ which is closely related to the vdW interactions, we see from Table I that the standard PBE functional predicts a value of 3.112 Å, which is larger than that measured experimentally by about 19%. Such bigger difference suggests that the vdW interactions can not be ignored when dealing with the physical properties of $Bi_2Te_3$ and similar structures. We further find that the calculation with optB86b-vdW functional gives a $d_{eqm}$ value of 2.704 Å, which is in better agreement with the experimental value of 2.612 Å. Among all the discussed vdW functionals, the ability of optB86b-vdW functional in predicting the lattice structure of $Bi_2Te_3$ is the best, and is thus used exclusively in our following calculations.

**B. Electronic properties of $Bi_2Te_3$ with PBE+optB86b-vdW functional**

The importance of appropriate treatment of vdW interactions lies not only in accurately predicting the structural parameters of $Bi_2Te_3$, but also the electronic properties. Fig. 2 displays the energy band structures of $Bi_2Te_3$ along several high symmetry lines in the irreduced Brillouin zone. For comparison, the calculations with standard PBE and PBE+optB86b-vdW are both shown. Indeed, we see the vdW has real effect on the energy band structure and obvious differences in the band shape and band gap are observed. The band gaps calculated with and without vdW interactions are 134 and 85 meV, respectively. It seems that the former agrees well with the experimentally measured gap of $150 \pm 20$ meV [33, 34, 35]. However, such coincidence is actually questionable since it is well known that DFT usually underestimates the band gap seriously. A careful search in the Brillouin zone using very dense $k$ mesh indicates that the CBM and VBM actually do not locate at those high-symmetry lines shown in Fig. 2. For the PBE+optB86b-vdW, the VBM and CBM appear at $k$ points (0.553, 0.395, 0.395) and (0.658, 0.553, 0.553) of the Brillouin zone. The energy difference between these two points should be regarded as the real band gap, which is calculated to be 108 meV. Our calculated results are summarized in Table II together with previous theoretical findings. It should be mentioned that the band extrema we obtained are very close to those found in Ref. 36 and Ref. 37. Note in some of these works [11, 36, 37, 38, 41], the lattice parameters used for band structure calculations are directly taken from the experimental data.

**C. Electronic properties of Bi2Te3 with PBE+optB86b-vdW+GW**

As mentioned above, the band gap of $Bi_2Te_3$ calculated with PBE+optB86b-vdW is still smaller than the experimental value. This is a known limitation of DFT which can be essentially solved by performing the GW calculations. The GW method could accurately predict the excited-state properties, such as the band gaps and quasiparticle energies, although more computational efforts are required. Fig. 3 shows the GW calculated energy band structures of $Bi_2Te_3$. Note here the vdW functional in the form of optB86b is by default included in the calculations. For comparison, the result without GW is also shown. There are some noticeable differences between these two band structures. First, in the case of standard DFT calculations, the VBM and CBM observed from the band structure (not necessarily the real ones) are both located in the Z-F direction; while the GW predicted VBM and CBM appear in the Z-F and Γ-Z directions, respectively. As a result, the gap between them increases from 134 meV to 197 meV when the GW correction is applied. Second, the DFT calculated band gap at the Γ point (0.49 eV) is much smaller than that of the GW calculation (0.68 eV). Third, the *camelback* shape of the highest valance band around the Γ point turns to be flattened when the GW approximation is taken into account. Finally, the GW tends to downshift the valance bands but upshift the conduction bands. Similar observations were also found in previous works [13, 14] focusing on the topological insulating properties of $Bi_2Te_3$.

It should be emphasized that the VBM and CBM of $Bi_2Te_3$ obtained directly from the energy band structure may not be the global band extrema. We thus perform an entire search of the Brillouin zone and the real VBM and CBM are found to be located at *k* points of (0.639, 0.560, 0.560) and (0.658, 0.579, 0.579), respectively. The real energy gap with GW corrections is thus calculated to be 157 meV, which is in excellent agreement with the experimental value [33, 34, 35]. Note the locations of VBM and CBM and the gap value in the present work are very close to those found in Ref. 11. Our GW calculated results are also summarized in Table II.

Up to now, we become aware of the importance of the vdW interactions and quasiparticle corrections. Only when both effects are appropriately taken into consideration can we give an accurate prediction of the structural and electronic properties of $Bi_2Te_3$. There is no doubt that such effects will also play an important role in determining the transport properties of $Bi_2Te_3$, which will be discussed in the following section.

**D. Transport properties of $Bi_2Te_3$ with GW approximation**

As a traditional good thermoelectric material, the electronic transport properties of $Bi_2Te_3$ have been widely investigated. By using the Boltzmann theory and relaxation time approximation, the Seebeck coefficient *S*, the electrical conductivity $\sigma$, and the electronic thermal conductivity $\kappa_e$ can be essentially derived [37, 38, 41]. However, most of such kinds of calculations are performed without considering the GW approximation, and there is usually an underestimation of the Seebeck coefficients. To address this problem, here the Boltzmann transport theory is combined with the maximally-localized Wannier function basis to interpolate the band structures and band velocities. To incorporate the temperature dependence of the band structures, we assume that the band shape does not change with temperature, but shift the valence or conduction bands which gives a temperature-dependent band gap $\Delta E_g = (0.157 - 1.08 \times 10^{-4} T)$ eV [33, 34, 39].

Pristine $Bi_2Te_3$ is known to exhibit *p*-type intrinsic conduction below room temperatures and the saturation hole concentration is found to be $1.1 \times 10^{19}$ cm$^{-3}$ [39]. In Table III, we list a series of experimentally measured hole concentrations at temperature ranging from 100 to 500 K [39]. At these particular concentrations, the calculated in-plane Seebeck coefficient *S* as a function of temperature is plotted in Fig. 4(a). The experimentally measured values [39] are also shown as a comparison. For *p*-type conduction, we see that the Seebeck coefficients calculated with GW involved agree well with the experimental results below 300 K. At temperature above 300 K, however, the GW calculated results are much larger than those found experimentally. The observed rapid decrease of Seebeck coefficients can be explained as that more *n*-type carriers are excited at higher temperature. To clarify this point, we have also calculated the Seebeck coefficients for *n*-type carriers. Indeed, we see from Fig. 4(a) that the *n*-type Seebeck coefficients are almost negligible below 300 K and the major carriers remain holes, but are comparable to those of *p*-type at elevated temperature. If we consider both kinds of carriers, the calculated Seebeck coefficients agree well with the experimental results in the whole temperature range from 100 K to 500 K. It is worth mentioning that the maximum Seebeck coefficient at 300 K we obtained is 323 µV/K, which is very close to the theoretical limit estimated previously (320 µV/K) [40].

The next transport coefficient we consider is the electrical conductivity. Within the Boltzmann theory, the electrical conductivity $\sigma$ can only be calculated with respect to the relaxation time $\tau$. In principle, the relaxation time can be obtained by fitting the experimentally measured electrical conductivity or resistivity [39]. Table III lists all the fitted results. At room temperature, we see that the relaxation time is $1.06 \times 10^{-14}$ s which is consistent with previously assumed values [41, 42]. In Fig. 4 (b), we plot the calculated power factors ($S^2\sigma$) of $Bi_2Te_3$ along the in-plane direction with respect to temperature. We see again that the calculated results match well with those measured experimentally below 300 K. If both the major and minor carriers are considered (especially at higher temperature), we find good agreement between our theoretical prediction and the experimental data.

For the calculations of electronic thermal conductivity $\kappa_e$, we use the Wiedemann-Franz law $\kappa_e=L\sigma T$, where $\sigma$ is the electrical conductivity, $T$ is the temperature, and $L$ is the Lorentz number. For traditional metals, the Lorentz number is $2.45 \times 10^{-8}$ V$^2$/K$^2$. However, for most semiconducting thermoelectric materials, it is lower than this value. Here we use a Lorentz number of $1.54 \times 10^{-8}$ V$^2$/K$^2$ as determined previously for bulk $Bi_2Te_3$ [43, 44]. With all the transport coefficients available, we can now evaluate the thermoelectric performance of $Bi_2Te_3$ by calculating the figure of merit, or $ZT=S^2\sigma T/(\kappa_e+\kappa_l)$. Note here the experimentally measured lattice thermal conductivity $\kappa_l$ is inserted [39]. Fig. 4 (c) shows the calculated *ZT* value of $Bi_2Te_3$ along the in-plane direction as a function of temperature. The experimentally measured results [39] are also shown. We do not plot the data above 400 K since the corresponding lattice thermal conductivity measured in Ref. 39 is absent. In the whole temperature ranges considered, we see the calculated temperature dependence of *ZT* value exhibits very good agreement with that measured experimentally. The coincidence of our theoretical calculations with the experimental data emphasizes the importance and accuracy of vdW interactions and GW approximation in predicting the electronic and transport properties of $Bi_2Te_3$.

## IV. SUMMARY

In summary, we have demonstrated the important role of vdW interactions and quasiparticle corrections on the electronic and transport properties of $Bi_2Te_3$, which are usually ignored in the previous DFT calculations concerning its thermoelectric performance. We find that using the exchange correlation energy in the form of PBE and the vdW functional in the form of optB86b predicts the lattice parameters very close to the experimental values. However, such approach gives an energy band gap of 108 meV, which is still smaller than that measured experimentally. To address such underestimation, we perform the GW calculations on the band structure of $Bi_2Te_3$ and make a fine search in the whole Brillouin zone to locate the global VBM and CBM. The GW calculated band gap is 157 meV and matches the experimentally result very well. Moreover, using the GW calculations and Boltzmann theory, the temperature dependence of the Seebeck coefficients, the power factor, and the *ZT* value are derived and all of them well reproduce the experimental data. Our theoretical approach highlights the strong capability and necessity of considering both the vdW interactions and GW approximation in accurately predicting the thermoelectric transport properties of $Bi_2Te_3$ and other similar systems such as SnSe, black phosphorous, and transition-metal dichalcogenide.

**ACKNOWLEDGEMENTS**

We thank financial support from the National Natural Science Foundation (Grant No. 51172167 and J1210061) and the "973 Program" of China (Grant No. 2013CB632502).

**Table I** Structural parameters of bulk $Bi_2Te_3$ calculated with various kinds of vdW functionals. Here $a_0$ and $\alpha$ correspond to the rhombohedral cell, while $a$ and $c$ correspond to hexagonal cell. $u$ and $v$ are the internal coordinates of Bi and Te1 atoms in terms of $(\pm u, \pm u, \pm u)$ and $(\pm v, \pm v, \pm v)$, respectively. $d_{eqm}$ is the equilibrium inter-QL distance. The experimental results from Ref. 26 are also listed for comparison.

|  | Exp. | PBE | vdW-D2 | optB86b-vdW | optB88-vdW | optPBE-vdW | vdW-DF2-rPW86 |
|---|---|---|---|---|---|---|---|
| $a_0$ (Å) | 10.476 | 10.970 | 10.747 | 10.570 | 10.697 | 10.889 | 11.148 |
| $\alpha$ (°) | 24.166 | 23.351 | 23.226 | 24.075 | 23.975 | 24.650 | 23.866 |
| $u$ (Bi) | 0.4000 | 0.3974 | 0.3984 | 0.3995 | 0.3992 | 0.3984 | 0.3988 |
| $v$ (Te) | 0.2095 | 0.2153 | 0.2122 | 0.2106 | 0.2118 | 0.2135 | 0.2141 |
| $a$ (Å) | 4.386 | 4.440 | 4.327 | 4.409 | 4.443 | 4.463 | 4.610 |
| $c$ (Å) | 30.497 | 31.998 | 31.357 | 30.778 | 31.153 | 31.741 | 32.477 |
| $d_{eqm}$ (Å) | 2.612 | 3.112 | 2.855 | 2.704 | 2.812 | 2.973 | 3.081 |

**Table II** The locations of VBM and CBM as well as the corresponding band gap of bulk $Bi_2Te_3$, calculated with and without GW approximations. The results from previous calculations are also listed.

| Reference without GW | VBM | CBM | gap (eV) |
|---|---|---|---|
| Youn and Freeman (Ref. 36) | (0.546, 0.383, 0.383) | (0.663, 0.568, 0.568) | 0.06 |
| Larson (Ref. 38) | (0.546, 0.383, 0.383) | (0.381, 0.500, 0.500) | 0.05 |
| Scheidemantel et al. (Ref. 41) | (0.652, 0.579, 0.579) | (0.663, 0.568, 0.568) | 0.11 |
| Kim et al. (Ref. 37) | (0.646, 0.549, 0.549) | (0.555, 0.397, 0.394) | 0.154 |
| Wang and Cagin (Ref. 30) | (0.662, 0.584, 0.584) | (0.673, 0.579, 0.579) | 0.049 |
| Kioupakis et al. (Ref. 11) | (0.37, 0.54, 0.37) | (0.58, 0.68, 0.58) | 0.087 |
| This study | (0.553, 0.395, 0.395) | (0.658, 0.553, 0.553) | 0.108 |
| Reference with GW | VBM | CBM | gap (eV) |
| Kioupakis et al. (Ref. 11) | (0.66, 0.58, 0.58) | (0.67, 0.58, 0.58) | 0.165 |
| This study | (0.639, 0.560, 0.560) | (0.658, 0.579, 0.579) | 0.157 |

**Table III** The fitted relaxation time of bulk $Bi_2Te_3$ at a series of temperature. Here, $n_p$ and $\sigma_{exp}$ are respectively the hole concentration and electrical conductivity measured in Ref. 39. $\sigma/\tau$ is the calculated electrical conductivity with respect to the relaxation time $\tau$.

| Temperature (K) | 100 | 150 | 200 | 250 | 300 | 350 | 400 | 450 | 500 |
|---|---|---|---|---|---|---|---|---|---|
| $n_p$ ($10^{19}$/cm$^3$) | 1.1 | 1.1 | 1.1 | 1.1 | 1.1 | 1.6 | 1.7 | 2.0 | 2.5 |
| $\sigma_{exp}$ ($10^4 \Omega^{-1}$m$^{-1}$) | 38.5 | 22.5 | 12 | 7.5 | 5.0 | 4.9 | 5.0 | 5.5 | 7.5 |
| $\sigma/\tau$ ($10^3 \Omega^{-1}$m$^{-1}$fs$^{-1}$) | 3.77 | 4.17 | 4.55 | 4.73 | 4.73 | 5.82 | 6.36 | 7.13 | 7.54 |
| $\tau$ ($10^{-14}$ s) | 10.2 | 5.39 | 2.64 | 1.58 | 1.06 | 0.84 | 0.79 | 0.77 | 0.99 |

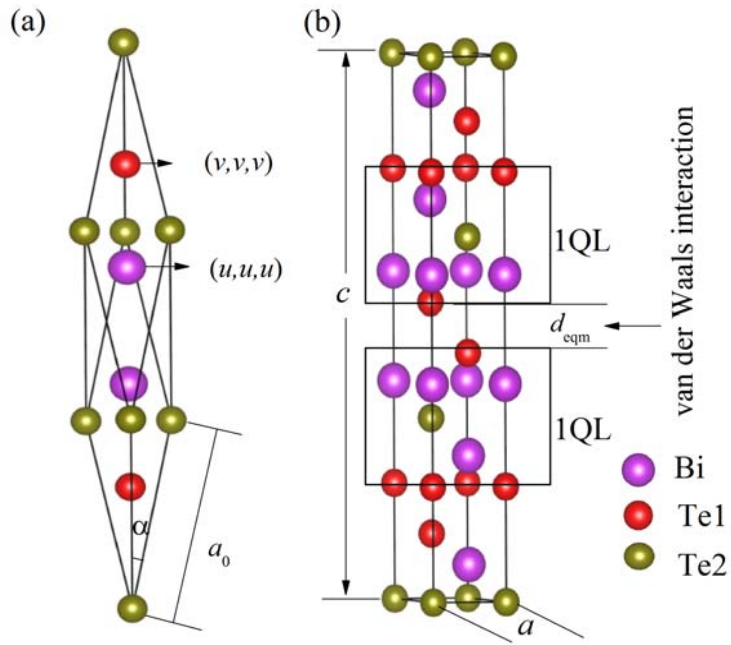

**Figure 1** The crystal structure of $Bi_2Te_3$ showing (a) rhombohedral unit cell, and (b) hexagonal unit cell.

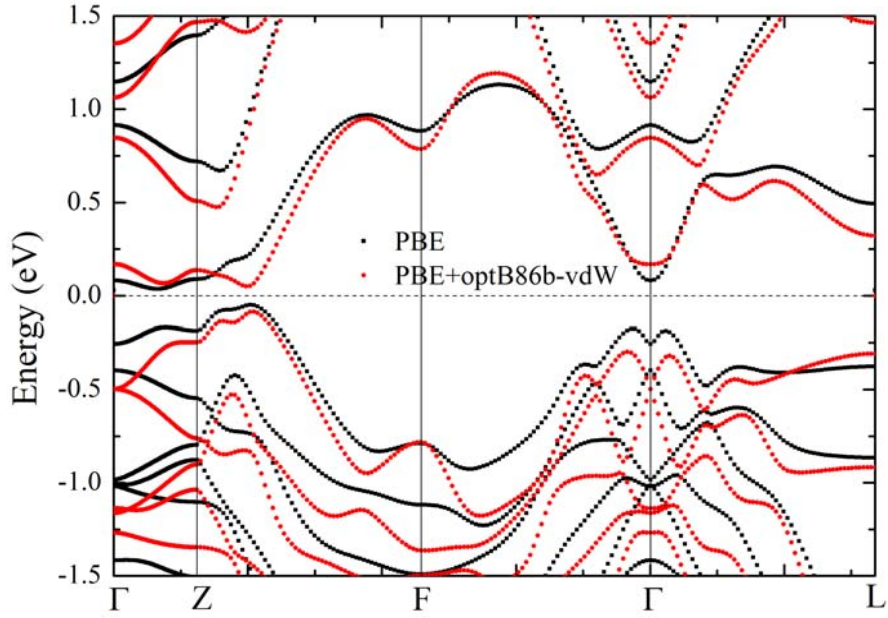

**Figure 2** Calculated energy band structures of $Bi_2Te_3$. The black and red lines correspond to the calculations with PBE and PBE+optB86b-vdW, respectively. The Fermi level is at 0 eV.

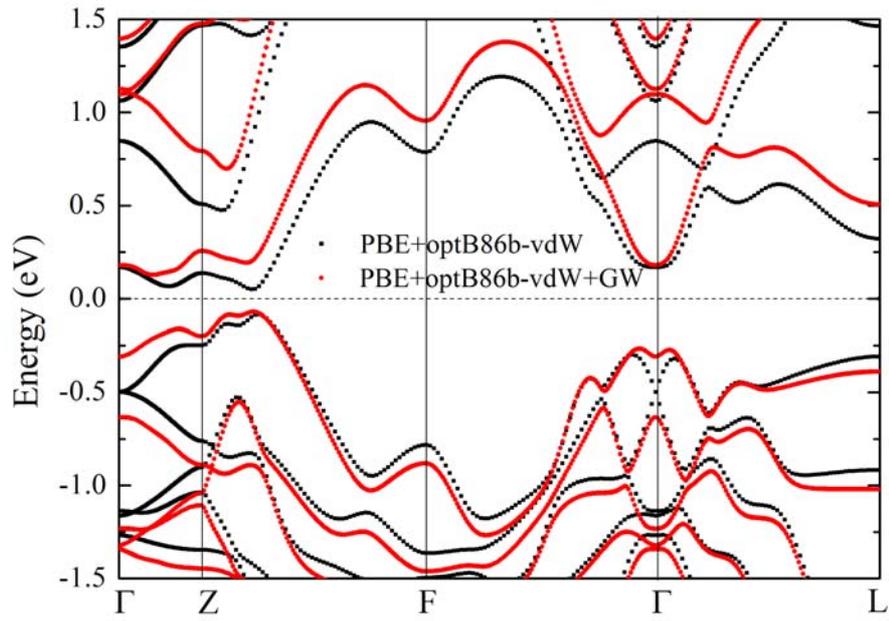

**Figure 3** Calculated energy band structures of $Bi_2Te_3$. The black and red lines correspond to the calculations with PBE+optB86b-vdW and PBE+optB86b-vdW+GW, respectively. The Fermi level is at 0 eV.

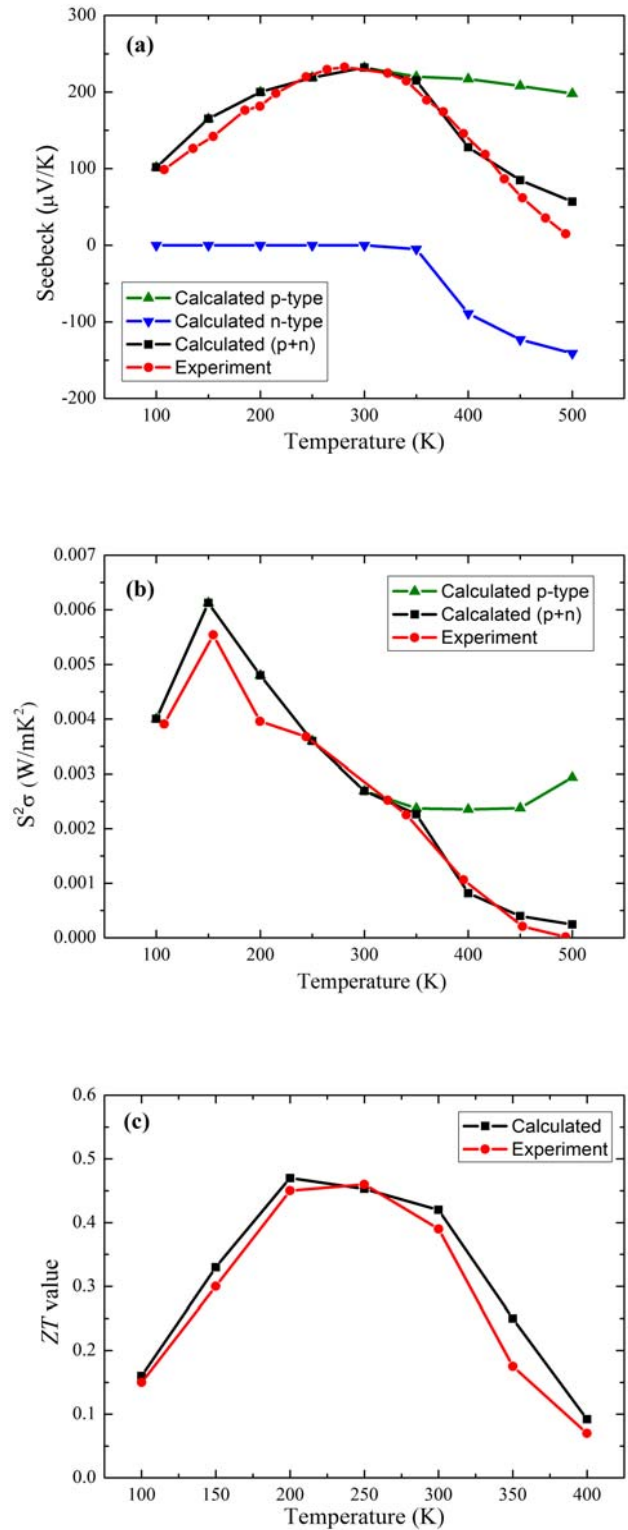

**Figure 4** Calculated temperature dependence of (a) Seebeck coefficient, (b) power factor, and (c) *ZT* value of $Bi_2Te_3$ at the experimentally measured carrier concentration. The experimental data from Ref. 39 are also shown for comparison.